\documentstyle[11pt,paspconf,epsfig]{article}

\begin{document}

\title{Search for new YSOs in Large Molecular Clouds and extinction mapping
using  DENIS and optical data}
\author{L. Cambr\'esy}
\affil{Observatoire de Paris, DESPA, 5 place Jules Janssen, F-92195 Meudon
Cedex, France}

\begin{abstract}
I present new results on the extinction and on the YSO content 
of the Chamaeleon I cloud  using the  star count method that I have implemented 
and applied to DENIS and optical data. 
Application of this method to other nearby Large Molecular Clouds is in
progress and is briefly outlined.  

\end{abstract}

\section{Introduction}
I recently reported on an original method to produce accurate extinction 
maps of large nearby molecular clouds. This method, based on adaptive star 
counts and wavelet decomposition, has been successfully applied to the nearby  
Chamaeleon I molecular cloud using massive $J$--band star counts provided by
DENIS (Cambr\'esy et al. 1997, Cambr\'esy 1998).
The combination of an extinction map and multicolour photometric 
near--infrared data such as those provided by DENIS, are powerful tools to 
investigate the young stellar populations of a star--forming region. In this
poster I present, on one side, new identifications of YSO candidates in Cha I
and, on the other side, further applications of my method to optical data and
their use to draw out extensive maps of moderate extinction.

\section{New YSO candidates in the Chamaeleon I cloud}
The detection of an infrared excess in a star, after removing the effect of  
cloud reddening, is a classical, but powerful method to discover new YSO 
candidates within a dark cloud. Every star detected in the Cha I cloud 
by DENIS has been dereddened using the extinction map previously drawn out.
These values of extinction must be considered as an upper limit of the
extinction suffered by the stars.
To select new YSO candidates, I have represented all stars in a 
colour--magnitude diagram ($K_s$ versus $J-K_s$) and picked up stars which
are off the main sequence by a distance corresponding to more than 8 
magnitudes of visual extinction. Fifty four new YSO candidates have been
selected in this way (Cambr\'esy et al 1998).

These new candidates are clearly concentrated in regions where the visual 
extinction is greater than 2. This argues in favour of their youth. 
Extinction could be underestimated if small size globules, not resolved in the 
extinction map (i.e. smaller than approximately 2 arcminutes) would lie on the 
line of sight of the object.
 
Assuming that the new candidates are mainly T Tauri stars, I can trace, 
together with already known T Tauri stars their luminosity function, and
attempt to derive their basic properties such as their mass and age.
Applying theoretical evolutionary tracks of  D'Antona \& Mazzitelli (1994)
and assuming the Miller and Scalo initial mass function, I can derive 
theoretical $K_s$ luminosity functions (KLF), $\phi_i(K_s)$, for different
ages ranging from $10^5$ to $10^8$ years.
I applied a singular value decomposition in order to solve the following
system of equations~:
$$
\sum_{i=10^5 yr}^{10^8 yr}{a_i \times \phi_i(K_s)} = \Phi(K_s)
$$
\noindent
where $\Phi(K_s)$ is the observed KLF.

\begin{figure}
\epsfig{figure=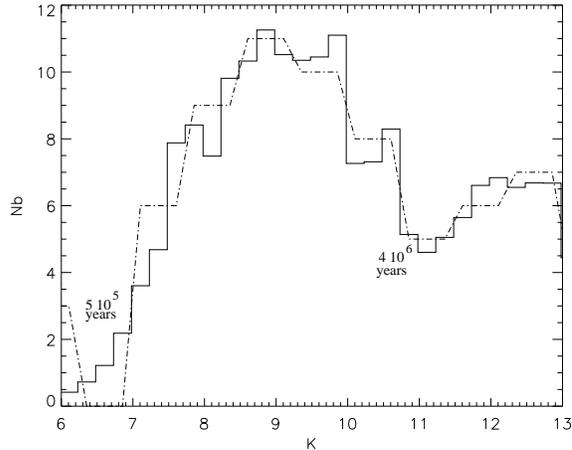,width=7.5cm}
\caption{The DENIS KLF (dashed line) and the fitted KLF (solid line)}
\label{KLF}
\end{figure}

Figure \ref{KLF} represents the observed  and the best fitted theoretical   
KLFs. This best fitting method suggests that two critical ages can be derived.
One, corresponding to $5\,10^5$ years, is consistent with the end of an active
star formation period in the cloud. The other, $4\,10^6$ years, represents the
maximum lifetime of the disc, since I assume that the criterion of infrared
excess used to select new candidates yields only stars surrounded by massive
circumstellar discs. 

\section{Extinction mapping with optical data}
\begin{figure}
\epsfig{figure=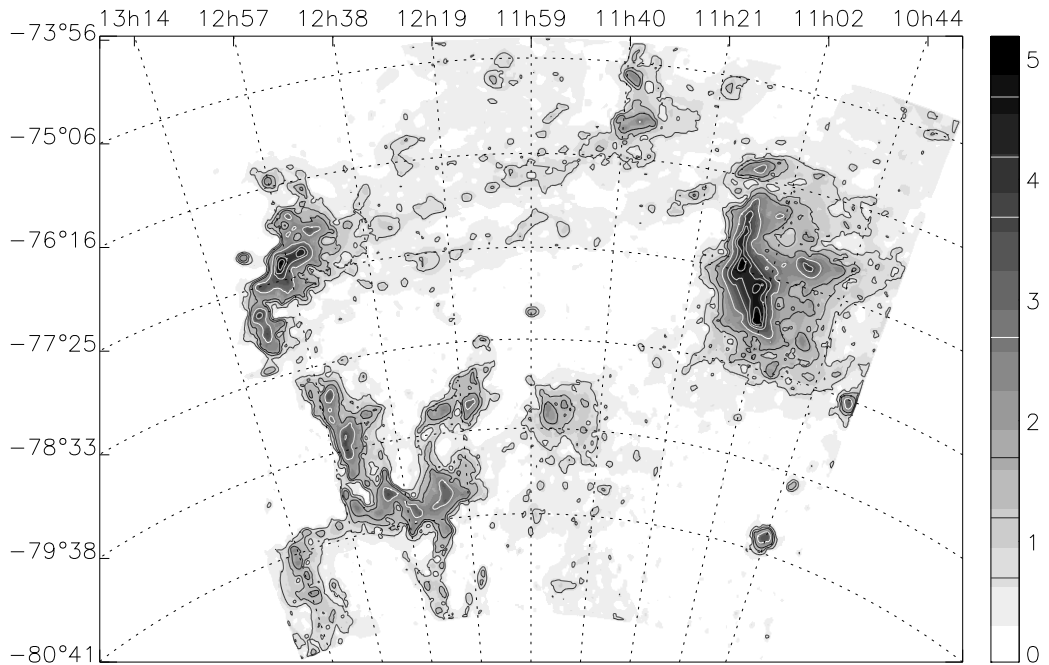,width=13.4cm}
\caption{Extinction map with iso--extinction contours of the Chamaeleon 
complex derived from optical star counts (USNO $B$)}
\label{cha}
\end{figure}

\begin{figure}[ht]
\begin{tabular}{ll}
\epsfig{figure=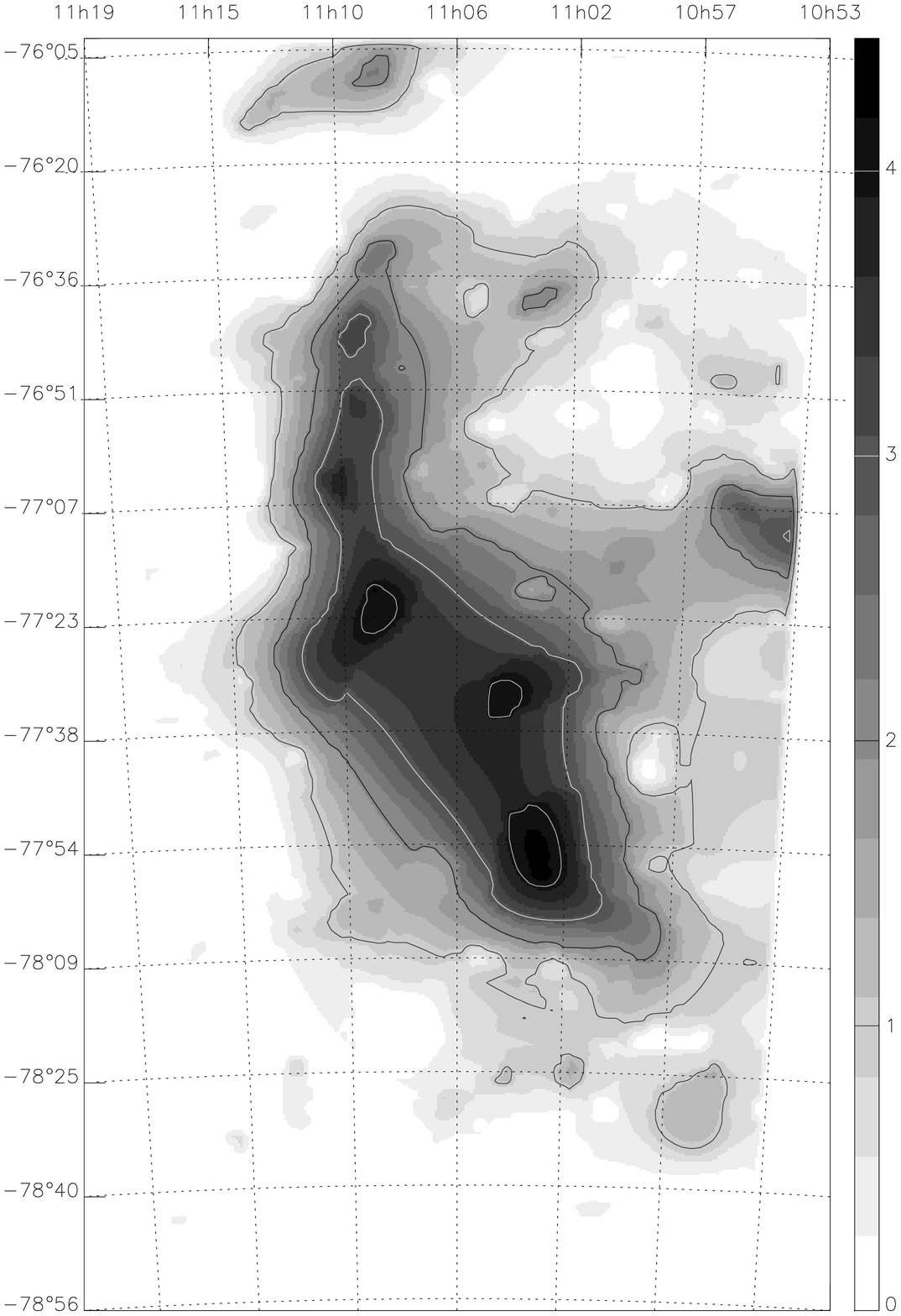,width=6.3cm} &
\epsfig{figure=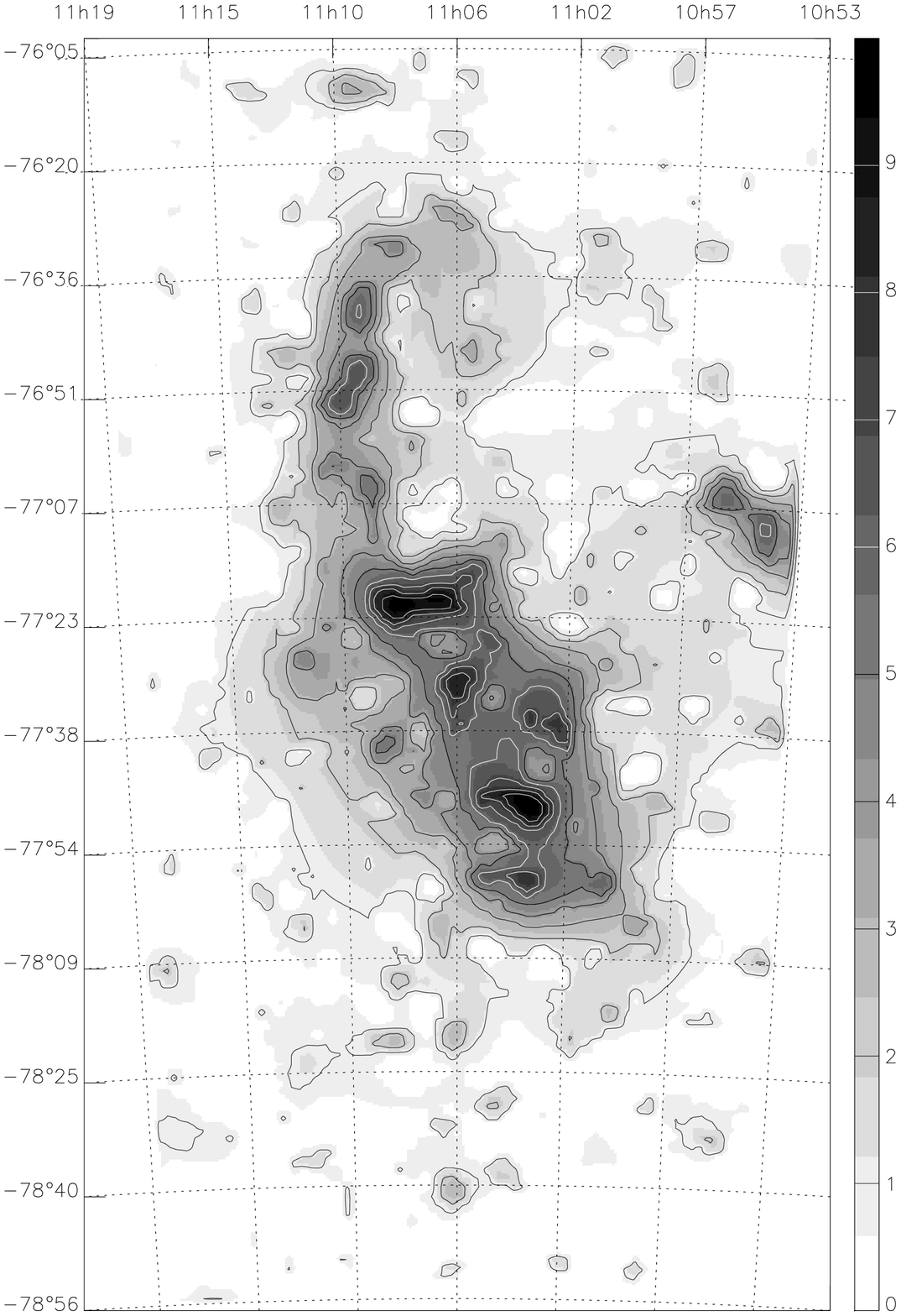,width=6.3cm} \\
\end{tabular}
\caption{Extinction map with iso--extinction contours of the Chamaeleon I cloud
derived from USNO $B$ star counts (left) and DENIS $J$ star counts (right)}
\label{chaBJ}
\end{figure}

The method that I have developed to produce maps of the extinction can be
applied to optical data provided by the digitisation of Schmidt plates.
The USNO catalogue (Monet 1996) provides $B$ and $R$ photometry for the all 
sky. These data allow large scale mapping of (moderate) extinction 
and are useful to identify the most obscured areas where near--infrared data
are required to provide further information, basically when the optical
extinction is greater than 5 magnitudes. Figure \ref{cha} and \ref{chaBJ}
represents the extinction map of the whole Chamaeleon complex ($\sim 80$
square degrees) and the comparison between $B$ and $J$ extinction maps of 
the Chamaeleon I cloud, respectively.
Comparison of the Chamaeleon map derived from $B$  and $J$ star counts 
(Cambr\'esy et al 1997) shows significant differences. Optical data indicate
a maximum of visual extinction of about 5 magnitudes, while a maximum of 
10 A$_V$ is reached when using $J$ counts. Moreover 4 different cores
with A$_V$ greater than 7 can be separated thanks to $J$ counts.
Since the extinction in $J$ band is only one third of the visual extinction,
near--infrared data allow deeper investigations of the cloud. 
For areas with low extinction, optical data are the most useful. The large
scale map of the Chamaeleon complex shows the possible dust connection 
between the different components of the whole cloud.

\section{Conclusion}
Large scale multi--wavelengths optical and near--infrared digitised surveys 
provide immense databases that contain unprecedented numbers of stars with
good photometry. They provide the observational material that dramatically
renew the interest of the classical, but powerful {\em star count} method to
derive interstellar extinction. I have implemented new computer tools that
allow to take maximum benefit of these huge databases to draw out arcminute
resolution map of extinction and applied them to the freshly recorded DENIS
data on the Cha I cloud. These maps are very useful to identify intrinsically
red objects which are likely to be YSOs and T Tauri stars. I have started to
extend this method to most of the large molecular clouds such as the all
Chamaeleon complex, Carina Nebula, Coalsack, Corona Australis, Lupus,
Musca, Orion, $\rho$ Ophiuchus, Rosetta, Taurus, Serpens, etc... 
using the incoming DENIS data, and, at larger scale, the optical data for
moderate absorptions. 
I expect that the combination of extinction maps based on B, R, I, J, K counts
will strongly improve our knowledge of the distribution and physical
properties of the dust within clouds and in the Milky Way, in general.

\end{document}